\documentclass[10pt,conference]{IEEEtran}
\usepackage{graphicx}
\usepackage{amsmath}
\usepackage{url}

\usepackage{xcolor}

\begin{document}
%
\title{Rethinking Software Development as a Self-Adaptive Socio-Technical System \\
}


\author{\IEEEauthorblockN{Javier C\'amara}
\IEEEauthorblockA{ITIS Software, Universidad de M\'alaga\\
jcamara@uma.es}\\

}


%


\maketitle

\begin{abstract}

Agentic AI is substantially shortening the timescale at which development context (e.g., goals, evidence, uncertainty, and risk) evolves. However, human--agent organization across development tasks is typically treated as fixed, which is problematic because development configurations may become suboptimal as the development context changes. We propose viewing software development as a \emph{self-adaptive socio-technical system} in which both the software project and the development configuration (participants, responsibilities, authority, information, and verification) adapt as the development context evolves. This yields two coupled forms of adaptation: evolving the software and reconfiguring how subsequent engineering is performed. We illustrate the idea with  a proof of concept, outlining challenges and  future plans. 
\end{abstract}

%
\IEEEpeerreviewmaketitle

\section{Introduction}
\label{sec:introduction}

Software development has traditionally been organized around relatively
stable assumptions about \emph{who} performs engineering work and
\emph{how} that work is coordinated. Even when responsibilities shift during
a project, the development configuration (which participants are involved,
what authority they have, what information they receive, and how their work is
verified) is typically established and revised through human coordination.
This is reasonable while changes in the development situation occur slowly
enough for developers and managers to recognize them, assess their
implications, and reorganize the work accordingly.

Agentic AI challenges this assumption. Agents can now perform extended
sequences of engineering activities across requirements, implementation,
verification, and evolution, substantially shortening the timescale at which
development state changes and new evidence becomes available. Recent evidence
on AI-assisted development similarly suggests that increased development
speed requires correspondingly faster feedback and control
mechanisms~\cite{DORA2025}. As engineering proceeds, the configuration that
was appropriate moments earlier may cease to be appropriate: ambiguity may be
discovered, risk may increase, an agent may repeatedly fail, independent
verification may become necessary, or human understanding of a critical
change may need to be restored. Conversely, continued human involvement may
be unnecessary once uncertainty or risk has subsided. In general, a fixed configuration cannot remain optimal as the development context evolves.

The resulting challenge is not only to determine \emph{which} development
configuration is preferable, but to do so at the pace at which agentic
development evolves. Continuously tracking the relevant state, comparing
alternative configurations, and reassigning responsibilities in a timely
manner places a growing coordination burden on human developers. This creates
a timescale mismatch between increasingly fast development activity and
human-mediated process adaptation. The situation echoes the motivation for
autonomic computing and self-adaptive systems: as computing systems became too complex and dynamic for
administrators to manage directly, management decisions were moved into
automated feedback loops operating under high-level human
objectives~\cite{KephartChess2003,de2013software}. Agentic AI now creates a parallel situation at the level of the \emph{software development process} itself.

To tackle this challenge, we propose rethinking software development as a \emph{self-adaptive socio-technical system} in which heterogeneous human and artificial participants collaboratively evolve a software project through feedback. Under this perspective, adaptation concerns not only the software artifacts being produced (e.g., requirements, designs, code, tests), but also the development system itself. As evidence accumulates, tasks may be redistributed, specialized agents recruited or dismissed, human participation requested, agent authority increased or constrained, or verification effort strengthened. This view builds on the insight that humans can be integral participants in adaptive systems, contributing information, decisions, and actions whose usefulness depends on their capabilities and circumstances~\cite{gil2016designing,yigitbas2021enhancing,CamaraEtAl2015,CalinescuEtAl2019,evers2014user}. The crucial shift in our proposal is to make the \emph{configuration of the human--agent development system itself} an explicit target of adaptation.

This shift is important because the best development configuration is not necessarily the one that maximizes immediate automation or throughput. Consider a change that agents can implement and verify autonomously with high confidence. Completing it without human participation may minimize time and financial cost. However, repeated choices of this kind can leave the human team with progressively less understanding of how and why the software has evolved. Recent work characterizes this erosion of shared understanding as \emph{cognitive debt}, which AI-assisted development may accumulate even when the generated software is technically adequate~\cite{Storey2026}. Related evidence suggests that people may adopt AI-produced judgments with limited critical scrutiny, including when those judgments are incorrect~\cite{ShawNave2026}. Hence, deliberately involving a human may sometimes be preferable despite its immediate cost, just as increased verification, specialized expertise, or reduced agent authority may become worthwhile as uncertainty or risk changes. Hence, human involvement  becomes one possible tactic, among several, to adapt development configuration, rather than merely a fallback mechanism for agent failure.

This perspective is related to, but distinct from, emerging work on
agentic and multi-agent software engineering. Recent approaches distribute
software-engineering activities among autonomous or specialized agents,
coordinate multi-agent workflows, and increasingly consider human--agent
collaboration~\cite{HeEtAl2025,HassanEtAl2025,OtoumElkhalili2026}.
Recent visions also broaden agentic software engineering beyond coding toward
a whole-process, explicitly socio-technical view involving evolving human and
agent roles, activities, and artefacts~\cite{Hoda2026}, while complementary
work emphasizes trust, intent understanding, verification, and effective
human--agent collaboration as prerequisites for dependable agentic
development~\cite{Roychoudhury2025,RoychoudhuryEtAl2026}.
Our contribution is different. We treat the
\emph{development configuration itself} (which participants are involved,
their responsibilities and authority, the information available to them, and
how their work is verified) as a runtime adaptation target. The development
system can continuously reason from feedback about whether its current
human--agent organization remains appropriate and reconfigure it as development
state and objectives evolve. Such decisions are governed by
explicit trade-offs among development-level goals (e.g., speed, cost,
assurance, and preservation of human understanding) so that no single
configuration can be assumed to be optimal independently of the current context.

This paper makes three contributions. First, it conceptualizes software development as a self-adaptive socio-technical system whose state includes both an evolving software project and an evolving configuration of human and artificial participants. Second, it identifies the central feedback problem that follows from this view: continuously determining how participants, responsibilities, information, authority, and verification should be configured according to development state, objectives, evidence, uncertainty, and risk. Third, it outlines a lightweight proof of concept in which development configuration is treated as an explicit adaptation variable, and derives a research agenda around development-state awareness, participant modeling, and adaptive coordination. 

\section{Software Development as a Self-Adaptive
Socio-Technical System}
\label{sec:model}

\begin{figure}[t]
    \centering
    \includegraphics[width=\linewidth]{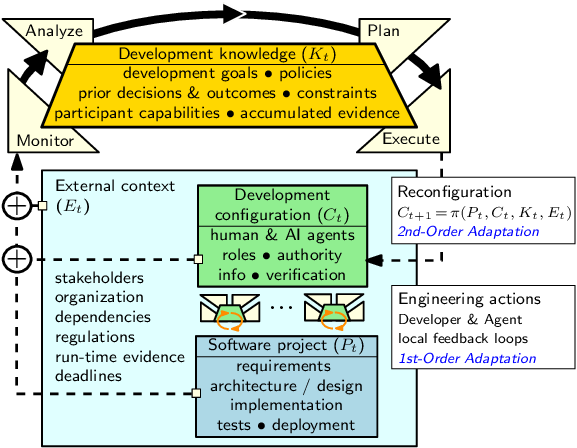}
    \caption{Notional representation of software development as a self-adaptive socio-technical system, illustrated on the MAPE-K paradigm~\cite{KephartChess2003}.
    }
    \label{fig:self-adaptive-dev-system}
\end{figure}

We conceptualize agentic software development as a socio-technical system in which human and artificial participants collaboratively evolve a software project through a feedback process  (Figure~\ref{fig:self-adaptive-dev-system}). The key difference from prevalent views of AI-assisted development is that we treat not only the software project, but also the configuration of the development system as something that can change while engineering proceeds.

\paragraph{Software project and development knowledge}
Let \(P_t\) denote the state of the software project at time \(t\). It captures what is being engineered, including product requirements, design and architectural decisions, implementation, verification artifacts, deployment artifacts, and evidence about the resulting system. For example, a requirement stating that users must be able to delete their personal information while legally required transaction records are retained belongs to \(P_t\), since it constrains the behavior of the software being built.

We distinguish project state from development knowledge \(K_t\), which captures information governing \emph{how} the project should be engineered. This includes development objectives, organizational policies and constraints, previous decisions and outcomes, accumulated evidence, and knowledge about the capabilities and limitations of human and artificial participants. Examples include staying within budget, requiring independent review for high-risk changes, or maintaining sufficient human understanding of a critical subsystem. 

Maintaining this knowledge is crucial in AI-intensive development.
If goals, constraints, or decision rationale are poorly externalized or
allowed to erode, the project can accumulate \emph{intent debt}, leaving
humans and agents without a reliable account of what the system is for or
why previous decisions were made~\cite{Storey2026}. In our model, such
degradation affects the information available in \(P_t\) and \(K_t\) and
may trigger adaptation to recover or clarify the missing intent.

\paragraph{Development configuration}
Engineering is performed at any point under a development configuration \(C_t\), which identifies, not only the participants currently assigned to a task, but the broader organization of development:

\vspace{0.07cm}
\centerline{ $C_t=\langle  \mathit{participants},
\mathit{roles}, \mathit{authority},
\mathit{info},\! \mathit{verification} \rangle$}
\vspace{0.05cm}

Participants may include developers, domain experts, other
human stakeholders, as well as artificial agents with various
capabilities and tool access. We draw on established feedback-loop
models for autonomic and self-adaptive systems, such as MAPE-K, which structures
adaptation around monitoring, analysis, planning, and execution over
shared knowledge~\cite{KephartChess2003}. Frameworks like Rainbow show how such
reasoning can be operationalized using monitored properties, runtime
models, adaptation strategies, and effectors~\cite{garlan2004rainbow}.
Prior work has also shown that humans can be modeled as integral
participants in adaptation, contributing information, decisions, or
actions whose suitability depends on factors such as capability and
context~\cite{gil2016designing,CamaraEtAl2015,CalinescuEtAl2019}. Our proposal
transfers these ideas to a different managed system: the
software-development system itself. The entities being reconfigured
are not primarily components of the running software, but the
human--agent organization through which software development is
performed.

\(C_t\) is not assumed to remain fixed. Evidence produced during development may indicate that a different configuration is better suited to the current situation. An AI agent may initially work autonomously, for example, but unresolved requirements ambiguity may warrant human domain input; inconsistent verification results may justify independent review; and insufficient human understanding may motivate a walkthrough even when agents can complete the task autonomously.

\paragraph{Two coupled forms of adaptation}
This perspective exposes two forms of adaptation. \emph{First-order adaptation} changes the software project through engineering actions:
\centerline{$P_t \longrightarrow P_{t+1}$}
Such actions may refine requirements, revise design decisions, or change implementation artifacts.

In contrast, \emph{second-order adaptation} changes the development configuration responsible for performing those actions:
\centerline{$C_t \longrightarrow C_{t+1}$}
For example, a development system may recruit a specialist, involve a human stakeholder, restrict an agent's authority, increase independent verification, or provide additional context. The resulting configuration then determines how subsequent first-order adaptation is performed.

The two forms are coupled through feedback. Engineering actions produce evidence about both the project and the development process; this evidence updates \(K_t\) and may reveal that the current \(C_t\) is no longer appropriate. Conceptually, the next development configuration can therefore be viewed as:

\centerline{$C_{t+1} = \pi(P_t,C_t,K_t,E_t)$}

We deliberately leave the realization of $\pi$ open. It could be
implemented through a MAPE-K-style feedback loop using rules,
planning, utility-based reasoning, learning, or combinations thereof. Formal reference models for self-adaptation such as FORMS~\cite{FORMS} and MORPH~\cite{MORPH} can be particularly useful for engineering coupled forms of adaptation because they illustrate how distinct forms of adaptation can be represented as
first-class concerns and coordinated when necessary.

The distinction is also important with respect to adaptive agent workflows.
Recent work optimizes agentic computation graphs by selecting, generating,
or revising which agents and components participate, their dependencies,
and information flows at or before run time~\cite{YueEtAl2026}.
Such mechanisms could provide valuable implementation techniques for the
adaptation function $\pi$. Our scope, however, is broader than workflow
optimization. A development configuration also captures human participation,
decision authority, verification responsibilities, and development-level
objectives whose effects may extend beyond successful execution of the
current task. Our perspective is concerned not only with optimizing an agent
workflow, but more broadly, with adapting the socio-technical system responsible for
developing and evolving the software.

\paragraph{Adaptation is multi-objective}
The preferred development configuration cannot generally be determined by maximizing automation or short-term productivity alone. For instance, involving a human in a change that agents could complete autonomously can increase completion time and financial cost while providing independent scrutiny and maintaining human understanding of the evolving software. Conversely, extensive human involvement in routine work may impose unnecessary delays and costs. Coordination is a multi-objective adaptation problem whose consequences can extend beyond the development action(s) currently being performed.

\section{A Lightweight Proof of Concept}
\label{sec:poc}

\paragraph{Scenario}
Consider a development task in which an existing online service must support
deletion of personal data while preserving transaction records that must be
retained. A general-purpose agent may initially explore the change
autonomously. If it encounters unresolved ambiguity in the retention rationale,
the development configuration $C_t$ can adapt by involving a human stakeholder
to clarify intent. If the task is subsequently assessed as high risk, a
specialist agent may be introduced for implementation. Verification may then
adapt from standard checks to independent verification when stronger assurance
is needed. Finally, even after technical verification succeeds, the system
may select a human walkthrough when preserving human understanding of a
critical change is an important development objective. Hence, these interventions are not fixed workflow stages, but context-dependent reconfigurations of
the participants, responsibilities, and verification arrangements used to
continue development.

\paragraph{Model}
We abstract these decisions as a Markov decision process (MDP) in
PRISM~\cite{Kwiatkowska2011PRISM}. The model follows four stages, namely
\textsc{Explore}, \textsc{Implement},
\textsc{Verify}, and \textsc{Review}. At each stage, the controller can
select between a lean configuration and a richer alternative: agent-only
exploration or human consultation; general-agent or specialist implementation;
standard or independent verification; and autonomous completion or a human
walkthrough. The state records unresolved intent ambiguity, task risk, latent
defects, verification evidence, and whether sufficient human understanding is
preserved. The review decision occurs after technical verification has passed,
allowing the model to represent human involvement whose purpose is not to
compensate for agent failure. 
The PRISM model captures only the decision-making core that could form
part of the analysis/planning functionality of a MAPE-K-style loop adapting
$C_t$~\cite{KephartChess2003,garlan2004rainbow}.

\paragraph{Evaluation}
All probabilities and resource costs are synthetic and encode only qualitative
relationships between development conditions and participant capabilities. We
analyse alternative configurations under representative development-level
priorities, including resource efficiency, assurance, and preservation of human
understanding, and perform stage-specific sensitivity analyses that vary
relevant state, capability, and objective parameters. The complete replication package is available at: \url{https://doi.org/10.5281/zenodo.22706335}.

\paragraph{Findings}
The results provide three observations. (1)~the preferable configuration
changes with the development situation even under fixed development objectives:
changes in intent ambiguity, risk, or participant capability can move the
decision between lean and richer configurations; (2)~ the same development
situation can favour different configurations when development-level priorities
change, illustrating how the objectives represented in $K_t$ can affect the
choice of $C_t$; (3)~adaptive coordination can outperform fixed
configurations when reconfiguration avoids the disadvantages of both
fixed extremes. This benefit is not universal: when resource efficiency
dominates, a fixed autonomous configuration may already be adequate. The review
analysis is particularly important for our argument, as it identifies regions
in which a human walkthrough is preferable even though technical verification
has succeeded and autonomous completion remains possible.

\paragraph{Interpretation and limitations}
Taken together, the results provide evidence for the decision relevance of
\emph{second-order adaptation}: the development system may need to adapt not
only the software being engineered, but also how humans and agents are
organised to engineer it. Our proof of concept is intentionally narrow and is
intended to establish feasibility rather than general effectiveness. The
adaptation space, development-state indicators, participant capabilities, and
decision model are simplified and hand-selected, and the running scenario does
not represent the diversity of real development settings. The numerical
parameters are synthetic and should not be interpreted as empirical estimates
of developer or agent performance. Moreover, longer-term outcomes such as
human understanding and cognitive debt are not directly measured; human
understanding is represented only through a simplified state variable.
Consequently, any observed benefit of adaptive coordination should be
interpreted as evidence that development configuration can be treated as a
meaningful adaptation variable, rather than as evidence that the particular
policy, parameterisation, or model employed is broadly optimal.

\section{Research Challenges}
\label{sec:challenges}

The proof of concept deliberately simplifies the information and reasoning required for adaptive coordination. Making the idea practical raises three connected research challenges.

\paragraph{Development-state awareness}
A self-adaptive development system must determine when its current configuration is no longer appropriate. Some relevant signals are directly observable, such as test failures, repeated agent attempts, tool errors, or disagreement among agents. Others are latent and harder to assess, including requirements ambiguity, inadequate context, rising risk, or erosion of human understanding. Research is needed to identify development-state properties relevant to coordination and reliable proxies for those that cannot be observed directly.

\paragraph{Participant modeling}
Effective reconfiguration also requires models of heterogeneous humans and artificial agents. Human participants may differ in expertise, availability, contextual knowledge, accountability, and cognitive load, while agents may differ in specialization, tool access, cost, latency, reliability, and permitted authority. Prior work on human involvement in self-adaptation demonstrates the usefulness of explicitly modeling factors such as capability, opportunity, and willingness~\cite{eskins2011multiple,CamaraEtAl2015,CalinescuEtAl2019}. Agentic development requires analogous evolving models for both human and artificial participants.

\paragraph{Adaptive coordination}
Given project state, development knowledge, context, and participant models, the system must determine whether and how to change \(C_t\). Candidate adaptations may modify participant composition, responsibilities, authority, information, interactions, or verification effort. These choices are inherently multi-objective: human review may increase time and cost while reducing risk or preserving understanding; additional agents may increase confidence while consuming additional resources. The core technical challenge is reasoning about such trade-offs under uncertainty and over time. Doing so requires evaluating whether reconfiguration is beneficial, looking beyond immediate task success. A configuration minimizing completion time may negatively affect quality, delivery stability, or the team's future ability to understand and evolve the software~\cite{DORA2025}. Conversely, locally inefficient actions may provide longer-term benefits, such as mitigating cognitive debt~\cite{Storey2026}. Development-system utility must therefore span immediate, medium-term, and longer-term outcomes~\cite{gheibi2022lifelong,gheibi2024dealing,zhu2024advancing}.


\section{Future Plans}
\label{sec:future}

We plan to explore the following research directions:

\paragraph{Richer development and participant models}
We will investigate richer representations of development state and of the human and artificial participants involved. In particular, we aim to study how development traces can be used to estimate properties difficult to observe directly, including uncertainty, risk, participant reliability, contextual knowledge, and human understanding, and how these estimates can evolve from evidence gathered during development.

\paragraph{Sequential adaptive decision-making}
Coordination decisions can have consequences beyond the activity in which they are made. We plan to investigate sequential, multi-objective decision mechanisms that reason about uncertainty and longer time horizons, balancing immediate concerns such as time and financial cost against quality, risk, verification, and preservation of engineering knowledge.

\paragraph{Empirical validation with human developers}
Finally, we will evaluate adaptive coordination on increasingly realistic tasks involving human developers and AI agents, comparing it with  fixed configurations and assessing both immediate outcomes and longer-term effects on developer understanding and ability to subsequently modify the software.

Jointly exploring these directions will enable us to determine when adapting the development configuration provides advantages over  statically orchestrated human--agent workflows, and which development objectives, signals, and decision mechanisms are necessary to attain such advantages.



\IEEEtriggeratref{13}


\bibliographystyle{IEEEtran}
\bibliography{biblio}

\begin{thebibliography}{1}

\bibitem{IEEEhowto:kopka}
H.~Kopka and P.~W. Daly, \emph{A Guide to \LaTeX}, 3rd~ed.\hskip 1em plus
  0.5em minus 0.4em\relax Harlow, England: Addison-Wesley, 1999.

\end{thebibliography}
%

\end{document}